\documentclass[aps,pra,twocolumn]{revtex4-1}
\usepackage{amssymb}
\usepackage{amsfonts}
\usepackage{amsthm}
\usepackage{amsmath}
\usepackage{graphicx}
\usepackage[titletoc]{appendix}
\newtheorem{protocol}{Protocol}

\def\ket{\rangle}

\begin{document}
\title{Quantum Private Comparison over noisy channels}
\author{Vikesh Siddhu}
\email[]{vsiddhu@andrew.cmu.edu} 
\affiliation{Department of Physical Sciences, Indian Institute of Science Education \& Research (IISER) Mohali,
Sector-81, SAS Nagar, Manauli P.O. 140306, Punjab, India}
\affiliation{Department of Physics, Carnegie Mellon University, Pittsburgh, Pennsylvania 15213, USA}
\author{Arvind}
\email[]{arvind@iisermohali.ac.in}
\affiliation{Department of Physical Sciences, Indian Institute of Science Education \& Research (IISER) Mohali,
Sector-81, SAS Nagar, Manauli P.O. 140306, Punjab, India.}

\begin{abstract}
Quantum Private Comparison~(QPC) allows us to
protect private information during its comparison.
In the past various three-party quantum protocols
have been proposed that claim to
work well under noisy conditions. Here we tackle
the problem of QPC under noise. We analyze the
EPR-based protocol under depolarizing noise, bit
flip and phase flip noise. We show how noise
affects the robustness of the EPR-based protocol.
We then present a straightforward protocol based
on CSS codes to perform QPC which is robust
against noise and secure under general attacks.

\keywords{Quantum cryptography \and 
Quantum private comparison \and Noisy channels \and CSS Code}

\end{abstract}

\maketitle

\section{Introduction}
\label{intro}
Quantum ideas have led to surprising developments in the
field of secure  communication. The most startling example is
that of cryptography, where quantum ideas have revolutionized
the field. While most classical cryptography schemes depend
on computational complexity for their security, quantum
cryptographic schemes~\cite{BB84,BB92,BBM92,E91} offer
security based on physical laws. There have been further
developments such as quantum secure direct
communication~\cite{SDC1,SDC2,SDC3}, quantum secret
sharing~\cite{QSS1,QSS2,QSS3}, quantum authentication and
quantum signatures~\cite{QIS1,QIS2,QIS3,QIS4}.

Secure multi-party computation allows several distrustful 
parties to jointly compute a function while keeping their 
inputs private~\cite{Yao1}, and is of fundamental 
importance in secure communication. A particular 
instance is to compute the equality function with just 
two parties~\cite{Yao1}. Quantum Private Comparison (QPC) aims to 
do the above computation without sharing the party's 
private information.  This is in contrast to 
quantum key distribution (QKD) which provides a
secure way to share private information.

Let Alice and Bob have private information $M_A$ and $M_B$
respectively.  QPC involves the computation of the function
$f(M_A,M_B)$ such that
\begin{equation} 
f(M_A,M_B) = 
\begin{cases}
0 & \text{if $M_A = M_B$}\\
1 & \text{if $M_A \neq M_B$}
\end{cases}
\end{equation} 
Furthermore, at the end of the protocol Alice and Bob do not
wish the other party to learn anything about their
information, apart from what can be inferred logically from
$f(M_A,M_B)$.  Lo~\cite{LO99} pointed out that the above
function $f(M_A,M_B)$ cannot be computed securely by two
parties alone. Hence a third party is needed to facilitate
the process. One might think that a three-party QPC is
trivial.  Both Alice and Bob can convey their information to
a trusted third party~(Charlie) and he can tell Alice and Bob the
outcome of the function $f$. The problem here is a little
different; Alice and Bob do not wish to disclose their
information to anyone, including Charlie and yet wish to compare
their private information. In fact, they do not want to transmit the
information at all. 
In the past several three-party quantum protocols have been
proposed~\cite{S1,S2,QPC,QPC1,QPCs}. They impose the following
restriction on the third party:

\begin{enumerate}
\item[(a)] Charlie tries to learn information about Alice and
Bob's input while being restricted to faithfully follow the
protocol. In other words he is semi-honest or \textit{honest but curious}.
\item[(b)] Charlie may know the positions at which $M_A$ and
$M_B$ differ, but not the actual bit values.
\end{enumerate}

Further, these protocols assume that all channels are noiseless
or remain silent on this aspect. We show that under the proposed 
restrictions, we can build a 
protocol to achieve QPC even under noisy conditions.
A slight modification of our protocol allows us to relax the condition,
that Charlie is honest. That is, he may not cooperate with Alice
and Bob and return False results. We also show how our protocol
is more efficient than similar quantum protocols~\cite{QPCs}.

It is hard to build perfect quantum channels and
hence we must build protocols that are robust
against noise. We choose a specific protocol
described by Tseng et. al.~\cite{QPC} and add noise
to its channels. We consider depolarizing noise,
bit flip and phase flip noise.  We show that the
protocol as such, is not robust under noise. We
note that three-party QPC involves transmission of
correlated keys between the parties, and that
under noise these correlations are altered.
Quantum error correction helps overcome the
effects of noise. We note that quantum error
correction and quantum cryptography have a deep
connection~\cite{PS}. Exploiting this connection,
we use the CSS quantum error correction
scheme~\cite{Steane} to transmit correlated keys
to relevant parties under noisy conditions in a
secure manner. This allows us to perform
three-party QPC under noisy conditions.
Further, by repeated use of our
protocol and through cooperation between Alice and
Bob, any dishonesty on the part of Charlie can
also be detected.

\section{EPR-based QPC protocol and noise}
\label{sec:QPC}
We review the
EPR-based  QPC protocol given in~\cite{QPC}. 
Alice and Bob have $n$ bit strings $M_A$ and $M_B$
respectively. They want to compare their information with the help
of a semi-honest third party called Charlie. Let Alice, Bob
and Charlie be connected by noiseless quantum channels that
can be eavesdropped upon and classical channels that can be
eavesdropped upon but not altered. For each qubit, we consider 
the computational
basis $\vert 0 \rangle $ and $\vert 1 \rangle$ and define
the rotated basis state as $\vert + \rangle=\frac{1}{\sqrt
2} (\vert 0 \rangle + \vert 1 \rangle)$ and 
$\vert - \rangle=\frac{1}{\sqrt
2} (\vert 0 \rangle - \vert 1 \rangle)$. For pairs of qubits 
the four Bell
states are defined as
\begin{eqnarray}
|\phi^{\pm}\ket = \frac{|00\ket \pm |11\ket}{\sqrt{2}}, \quad
|\psi^{\pm}\ket = \frac{|01\ket \pm |10\ket}{\sqrt{2}}.
\end{eqnarray}

Using these resources over the quantum channels and
classical communication over the 
classical channels, the secure QPC protocol proceeds as
follows:
\begin{protocol}
\end{protocol}
\begin{enumerate}
\item Charlie prepares a random $n$ bit string $C_T$. For
each bit of $C_T$ he prepares a quantum state. If the bit is
$0$ then he prepares one of the states from
$|\phi^{\pm}\ket$ (it does not matter which).  Otherwise, he
prepares one of the states from $|\psi^{\pm}\ket$.  Sequence
$T_A$ consists of the first half of each of these entangled
pairs, while $T_B$ consists of the second
halves.~\label{step1}
\item Charlie prepares two sets of decoys $D_A$ and $D_B$
randomly in the states: $|0\ket$, $|1\ket$, $|+\ket$ and
$|-\ket$. Charlie randomly interleaves $D_A$ with $T_A$  and
$D_B$ with $T_B$ to form $S_A$ and $S_B$, which are then
sent to Alice and Bob respectively. \label{step2}
	  
    \item Upon receipt of the complete sequences $S_A$ and
$S_B$, Alice and Bob signal Charlie to disclose the
positions and the basis ($\{|0 \ket, |1 \ket\}$ or $\{ |-
\ket, |+ \ket\}$) for measuring the decoys.\label{step3}
	  
    \item Alice and Bob measure the decoys in the
appropriate basis and consult over a classical channel to
check for eavesdroppers. If the error rate is more than a
predetermined rate then they abort the protocol, else they
proceed.\label{step4}
	  
    \item Alice and Bob measure the non-decoy particles in
the $Z$ basis to obtain bit strings $R_A$ and $R_B$
respectively. Note that each of $R_A$ and $R_B$ are
uniformly random while $R_A \oplus R_B = C_T$.\label{step5}
\item Alice and Bob calculate $C_A = M_A \oplus R_A$ and
$C_B = M_B \oplus R_B$.  They cooperate to calculate $C= C_A
\oplus C_B$ and send it to Charlie.\label{step6}
\item Charlie computes $R_c = C \oplus C_T$. $R_c$ has a
single non-zero entry if and only if $M_A \neq M_B$, in
which case Charlie outputs $1$, otherwise he outputs $0$.
\label{step7} 
\end{enumerate}
It is not hard to see that in the absence of noise and
eavesdropping, the protocol computes the function
$f(M_A,M_B)$ with certainty. We note that if an eavesdropper
(Eve) passes undetected then the output of the protocol can
be different from $f(M_A,M_B)$ because Eve can
tamper with the
non-decoy particles (she may cause $R_A \oplus R_B \neq
C_T$) and make the protocol malfunction.  It has been shown
that the above protocol is secure against certain insider
and outsider attacks~\cite{QPC} and hence computes
$f(M_A,M_B)$ with very high probability.

\subsection{One qubit noisy channels}
\label{sec:noise}
In the QPC protocol described above, perfect (noiseless)
single qubit quantum channels between Alice, Bob and Charlie
have been employed. In any real situation, noise can act
on these channels in a number of ways. Therefore, we need to
consider noisy one qubit channels instead of noiseless
channels and explore the possibility of carrying out QPC over
these noisy channels. We begin by describing  the noisy
channels and then figure out  their effect on the EPR-based 
QPC protocol.

The bit flip channel with
error probability $1-p$ is defined through its action on a
one qubit density operator $\rho$ via the action of the
bit flip gate $X$ as 
\begin{equation}\label{eq:xerror}
 \mathcal{F}(\rho) = (1-p) X \rho X^{\dag} + p  \rho.
\end{equation}
Similarly, 
the phase flip channel with error probability $1-p$ is
described through the action of the phase flip gate $Z$ as
\begin{equation}\label{eq:zerror}
 \mathcal{G}(\rho) = (1-p) Z \rho Z^{\dag} + p \rho. 
\end{equation}
The depolarizing channel with error probability $p$ is
\begin{equation}\label{eq:dpn1}
\mathcal{H}(\rho) = (1-p)\rho + 
\frac{p}{3}(X \rho X^{\dag} + Y \rho Y^{\dag} + Z \rho Z^{\dag}).
\end{equation}
The above equation admits the interpretation that the state
is acted upon by each Pauli operator with probability
$\frac{\displaystyle p}{\displaystyle 3}$ and remains 
unchanged with probability $1-p$. 

\subsection{QPC and depolarizing channels}
\label{sec:DP-QPC}
Let both the channels between Alice and Charlie (AC) and
between Bob and Charlie (BC) suffer from depolarizing noise.
If the error represented by the Pauli matrix $\sigma_A$ acts
on the AC channel and the error represented by $\sigma_B$
affects the BC channel then we call the combined error
$\sigma_A \sigma_B$.  From equation (\ref{eq:dpn1}) we see
that under depolarizing noise the channel acts such that
each Pauli matrix acts on the qubit with equal probability
$\frac{p}{3}$.  Since both the channels AC and BC are
independent the errors act independently.  Hence, the
probability for an $X_A X_B$ error is
$\frac{\displaystyle p}{\displaystyle 3} \cdot
\frac{\displaystyle p}{\displaystyle 3}$.  If an error acts 
such that it takes the state
$|\phi^{\pm}\ket$ to the state $|\psi^{\pm} \ket$ or
vice-versa then the protocol will return an incorrect
answer.  This happens because the flipping of a correlated
to an anti-correlated state and vice-versa, makes the string
$C_T$ an unfaithful record of the positions at which $R_A$
and $R_B$ differ. After the error has acted $C_T \neq C'_T$
where 
\begin{equation}\label{eq:ediff}
 C'_T \equiv R_A \oplus R_B
\end{equation}
So in step~\ref{step7} of Protocol~1, Charlie gets 
$R_c = (C_T \oplus C'_T) \oplus (M_A \oplus M_B)$ instead of
$R_c = M_A \oplus M_B$.

Under the action of depolarizing noise mentioned in equation
(\ref{eq:dpn1}) the probability that the state changes from
$|\phi^{\pm}\ket$ to $|\psi^{\pm}\ket$ or vice-versa is $r =
\frac{4p}{3}(1-\frac{2p}{3})$, which means that the
probability that $C_T$ and $C'_T$ differ at a given position
is $r$.  Even if there is a difference at a single position in
$C_T$ and $C'_T$ the protocol will give wrong results.
Let $n$ be the length of the strings and $P(C_T = C'_T)$ the
probability that $C_T = C'_T$. It is straightforward to see
that 
\begin{eqnarray}
P(C_T \neq C'_T) &= 1 - P(C_T = C'_T) \nonumber \\
 &= 1-(1-r)^n \label{E1}
\end{eqnarray}
Hence the protocol \cite{QPC} is not robust against any
amount of depolarizing noise.  For large $n$ and small $r$,
the error is linear in $r$.
\subsection{Bit and Phase Flip channels and QPC} 
\label{sec:BP-QPC}
Consider bit flip and phase flip noise in channels AC and
BC.  Suppose bit flip~(\ref{eq:xerror}) acts with
probability $p$ and phase flip~(\ref{eq:zerror}) with
probability $q$. The combined action of the error is given
by 
\begin{eqnarray}
\mathcal{F} \circ \mathcal{G}(\rho) &=& 
\mathcal{G} \circ \mathcal{F}(\rho) \nonumber \\   
&=& (1-q)pX\rho X + (1-p)qZ\rho Z + pq Y \rho Y \nonumber \\
&+& (1-q)(1-p)\rho. \label{eq:noisef}
\end{eqnarray}
Equation (\ref{eq:noisef}) gives the total action of noise on each channel.
Let the length of $C_T$ and $C'_T$ be $n$, then
\begin{equation}\label{E2}
P(C_T \neq C'_T) = 1-(1-2p(1-p))^n.
\end{equation}

Hence the protocol \cite{QPC} is robust against phase flip
but not bit flip noise.  For large $n$ and small $p$, the
error is linear in $p$.

We see that due to depolarizing noise and bit flip noise in
the communication channels between Alice~(Bob) and Charlie,
the protocol returns incorrect results. This is because
noise alters the quantum state being sent and consequently
the string $R_A$ and $R_B$. This alteration results in $C_T$
(the string with Charlie) becoming an  unfaithful record of
the correlations between $R_A$ and $R_B$.  In
general,
channels are noisy and any protocol fit for implementation
must be robust against noise. Hence we need to design
protocols that work even under noisy conditions.
\section{CSS Code based Protocol}
\label{sec:prot-CSS}
In order to perform three-party QPC under noise it is
necessary to preserve the information encoded in the quantum
states being sent by  Charlie to Alice~(Bob).  This will
ensure that $C_T$ remains a faithful record of the
correlations. One way to achieve this, is through error
correction  on the quantum states being sent to convey $R_A$
and $R_B$. We utilize CSS codes to perform error
correction~\cite{Steane}.  We note that these  codes have a
deep connection with QKD~\cite{PS}.

We propose a protocol for QPC that is robust under noise and
completely secure from attacks. The basic idea is to use the
CSS codes to securely transfer a known key from  Charlie
to Alice and Bob. This allows the QPC to work perfectly
under noise as long as the bit (phase) error rate is under
an acceptable limit.

\subsection{CSS Codes}
\label{sec:CSS}
We review the CSS codes~\cite{Steane,CS} and the protocol
for using CSS codes to perform a secure key distribution of
a known random key.

Suppose $C_1$ and $C_2$ are
$[n,k_1]$ and $[n,k_2]$ classical linear codes such that
$\{0\} \subset C_2 \subset C_1 \subset \mathbb{F}^{n}_2$,
$C_1$ and $C_2^T$ both correct $t$ errors. Then
$CSS(C_1,C_2)$ is an $[n, k_1 - k_2]$ quantum error
correcting code capable of correcting $t$ qubit errors.  For
$x \in C_1$ we define a code state \begin{equation} |x+C_2
\ket \equiv \frac{1}{\sqrt{|C_2|}}\sum_{y \in C_2} |x \oplus
y \ket \end{equation} where $\oplus$ is summation modulo
$2$. If $x,x'$ belong to the same coset in $C_2$ i.e. $x-x'
= y' \in C_2$ then they define the same code state, hence
the total number of distinct code states is the number of
cosets of $C_2$ in $C_1$, $|C_1|/|C_2| = 2^{k_1-k_2}$. Each
code state can be used to encode a distinct $n$ bit
classical string. This can then be exchanged between
interested parties.\newline The code state can get affected
by noise in the channel, which we must be able to correct.
It is sufficient to write the corrupted code state as
\begin{equation}\label{error:state}
\frac{1}{\sqrt{|C_2|}}\sum_{y \in C_2} (-1)^{(x+y).e_2}|x
\oplus y \oplus e_1 \ket \end{equation} 
where $e_1$ is the
$n$ bit string with a non-zero entry only at positions where
a bit flip has occurred and $e_2$ is a similar $n$ bit
string for phase flips. By correcting both these kind of
errors we can correct any kind of error~\cite{Steane,CS}.
In order to detect and correct errors we consider
$\sigma_{a(k)}$ the Pauli matrix acting on the $k^{th}$ bit,
where $a(k) \in \{x,y,z\}$. The operator $\sigma^{[l]}_{a}$
is defined as \begin{equation} \sigma^{[l]}_{a} =
\sigma^{l_1}_{a(1)} \oplus \sigma^{l_2}_{a(2)} \oplus
\cdots \oplus \sigma^{l_n}_{a(n)} \end{equation} $l$ is an
$n$ bit string and its $i^{th}$ entry is $l_i$ that takes
values from $\{0,1\}$.  By definition $\sigma^{0}_{a(k)} =
\mathbb{I}$. Note that eigenvalues of $\sigma_{a(k)}$ are
$\pm 1$.

In classical error correction if $F$ is a parity check
matrix for a code $M$, an error $y$ affecting the code word
$p$ giving $p'= p + y$ has syndrome $Fp' = Fy$ ($Fp=0$ by
definition). This syndrome is used to determine the most
likely error $y$. Note that the $m^{\text{th}}$ entry of the
column vector $Fy$ is $f_m \cdot p' \mod 2$, where $f_m$ is
the $m^{\text{th}}$ row in $F$.

For correcting the quantum state in Equation
(\ref{error:state}), we employ a measurement protocol along
similar lines. Let $H_1$ be the parity check matrix for
$C_1$ and $H_2$ for $C^{T}_2$~(the dual code of $C_2$). If
$l$ is the $i^{th}$ row of $H_1$ then we determine the
$i^{th}$ column entry for the bit flip error syndrome
$H_1\cdot e_1$ by measuring $\sigma^{[l]}_{z}$ with the
understanding that the eigenvalue $1 (-1)$ is mapped to $0
(1)$. Thus by measuring $\sigma^{[l]}_{z}$ for each row
$l \in H_1$ we obtain the full syndrome.  The
$i^{\text{th}}$ column entry for the phase flip error
syndrome $H_2 \cdot e_2$ is similarly obtained by measuring
$\sigma^{[l']}_{x}$ where $l'$ is the $i^{\text{th}}$ row of
$H_2$.  From these syndromes we can accurately get back
$e_1$ and $e_2$ using classical linear coding theory as long
as $wt(e_1) \leq t$ and $wt(e_2) \leq t$ respectively.  We
then correct the corrupted state and retrieve the encoded
state
\begin{equation}\label{cor1}
 \frac{1}{\sqrt{|C_2|}}\sum_{y \in C_2} |x \oplus y \ket 
\end{equation}

A generalized $CSS(C_1,C_2)$ code for any two $n$-bit
strings
$x$ and $z$ can be defined as  
\begin{equation}\label{eq:Q1}
|v+C_2 \ket \equiv \frac{1}{\sqrt{|C_2|}}\sum_{w \in C_2} 
(-1)^{z.w} |v \oplus x \oplus w \ket \quad v \in C_1
\end{equation}
We may use these code states.  Let $s\equiv(x,z)$ then we
denote the quantum code with the above code states as
$Q_{s}$.  For $x=0$ and $z=0$ $Q_{s}$ reduces to
$CSS(C_1,C_2)$.  If we measure $\sigma^{[l]}_{z}$($l \in
H_1$) and $\sigma_{x}^{l'}$ ($l' \in H_2$) on code state
(\ref{eq:Q1}) then we will obtain syndromes corresponding to
$H_1x$ and $H_2z$ respectively.  If there was a bit flip
error $e_1$ and a phase flip error $e_2$ on the code state
(\ref{eq:Q1}) then our syndrome measurements would be
corresponding to $H_1(x+e_1)$ and $H_2(z+e_2)$.  We can
recover the error with the understanding that we must
subtract $x$ and $z$ to retrieve the $e_1$ and $e_2$
respectively.  If we perform syndrome measurements on any
state $|\psi\ket$ and obtain that the syndrome are both null
vectors then we can conclude $|\psi\ket = |v+C_2 \ket \; v
\in C_1$ for some $v$.  The syndrome measurement projects
the state $|\psi\ket$ into the subspace spanned by $|v+C_2
\ket, \; v \in C_1$. Alternatively, if we obtain syndromes
corresponding to $H_1.x$ and $H_2.z$ for bit and phase flip
respectively, then we may conclude that $|\psi\ket$ has been
projected onto a subspace spanned by code states of
$Q_{s},\; s=(x,z)$. 
\subsection{The Protocol}
\label{sec:prot}
Let us first describe the CSS based protocol for sharing a
known randomly chosen secret key. Let us assume that a
secret key is to be distributed between Alice and Charlie.
\begin{protocol}\label{CSS}
\end{protocol}
\begin{enumerate}
\item Alice creates $n$ random check bits, a random $m$ bit
key $k$ and a random $2n$ bit string $b$.
\item Alice generates $s=(x,z)$ by choosing $n$-bit strings
$x$ and $z$ at random.
\item Alice encodes her key $k$ as $|k\ket$ using the CSS
code $Q_s$.
\item Alice chooses $n$ positions (out of $2n$) and puts
the check bits in these positions and the code bits in the
remaining positions.
\item Alice applies a Hadamard transform to those qubits
in those positions where $b$ is $1$.
\item Alice sends the resulting state to Charlie. He
acknowledges the receipt once he receives all qubits.
\item Alice announces $b$, the positions of the check
bits, the values of the check bits and the strings $s$.
\item Charlie performs Hadamard on the qubits where $b$ is
$1$.
\item Charlie checks whether too many of the check bits
have been corrupted, and aborts the protocol if so.
\item With the help of $s$, Charlie decodes the key bits
and uses them for the key.  \end{enumerate} The above
protocol works correctly and is unconditionally secure as
long as the noise is under a given threshold
value~\cite{PS}. The protocol for carrying out QPC under
noisy conditions is as follows

\begin{protocol}\label{CSS-QPC}
\end{protocol}
\begin{enumerate} 
\item
Charlie generates a random $n$ bit string $R_A$ and uses
the CSS Code based quantum error correction
protocol(Protocol~\ref{CSS})
to send it to Alice.
\item Charlie generates a random $n$ bit string $C_T$ and
computes $R_B = R_A \oplus C_T$
\item Charlie uses Protocol~\ref{CSS} to send $R_B$ to
Bob.
\item Alice and Bob compute $C_A = M_A \oplus
R_A$ and $C_B = R_B \oplus M_B$.
\item Alice and Bob collaborate together to compute $C =
C_A \oplus C_B$ and send it to Charlie over a public
channel.
\item Charlie computes $R_c = C \oplus C_T$. $R_c$ has a
single non-zero entry if and only if $M_A \neq M_B$, in
which case Charlie outputs $1$, otherwise he outputs $0$
\end{enumerate}
\begin{center}
\begin{figure}[h]
\centering
\includegraphics[scale =1.0]{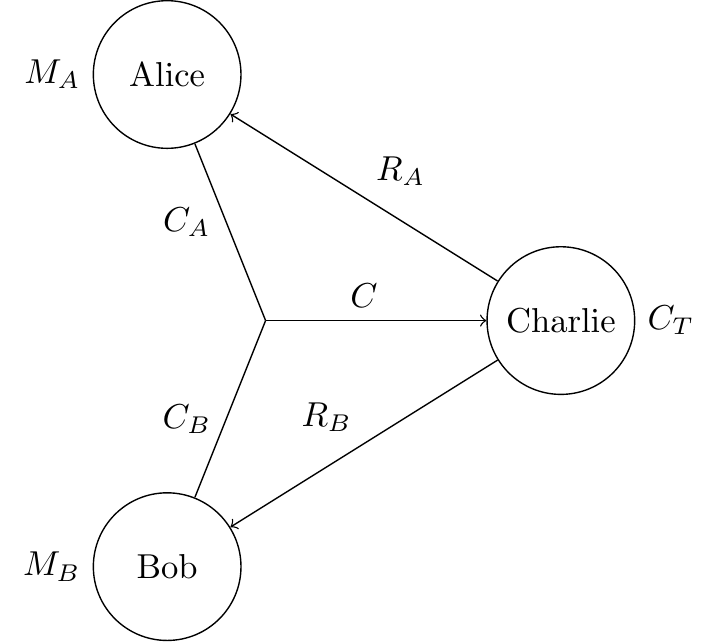}
\caption{The schematic diagram of the protocol where
Charlie generates random strings $R_A$ and $C_T$, using the 
CSS based protocol he sends $R_A$ to Alice and $R_B= R_A \oplus$
$C_T$ to Bob over the noisy channels. Alice and Bob encode their
respective messages $M_A$ and $M_B$ in $C_A$ and $C_B$. 
They collaborate to compute $C = C_A \oplus C_B$ and send 
it via a public channel to Charlie.}
\label{figure}
\end{figure}
\end{center}
The entire process is summarized in Figure~\ref{figure}.
It is easy to see that in the absence of noise and
eavesdropping the protocol computes the function
$f(M_A,M_B)$ correctly. In the presence of noise alone the
CSS based scheme can transmit keys correctly as long as
noise is within an acceptable level 
(the current acceptable
level of bit(phase) flip errors is 
$20.0 \%$~\cite{PhysRevA.66.060302,1176619}).  When
both noise and eavesdropping are allowed the protocol is
secure and gives correct results with very high probability.
We now show the security and correctness in the presence of
noise and eavesdropping.  We note that participant attacks are
stronger than non-participant attacks since participants always
have more information. We consider attacks by Alice and Bob to 
demonstrate the security of the protocol.

Consider an attack by Alice to gain information about $M_B$.
She
can attack the transmission channel between Bob
and Charlie,
and try to extract information by performing any physical
operation permitted by quantum mechanics. Alternatively she
may exploit side channel attacks which exploit loopholes in
the devices used to implement key distribution
~\cite{PhysRevLett.85.1330,PhysRevA.78.042333,nat_phot,Lamas-Linares:07,Lydersen:10,PhysRevA.75.032314,PhysRevLett.107.110501,trojan,time-shift}.
These two are fundamentally different kinds to attack.

Let us first analyze a direct attack on the transmission by
Alice.  She has access to $M_A$, $C_B$, $C_A$ and $R_A$. We
may assume that $M_A$ contains no information about $M_B$.
We note $M_B = R_B \oplus C_B$, hence information about
$R_B$ implies information about $M_B$ and vice-versa.  Alice
can gain information about $R_B$ through $C_T$ ($R_B = C_T
\oplus R_A$), alternatively she may intercept the
communication between Bob and Charlie.  The semi-honest
nature of Charlie ensures that Alice does not learn anything
about $C_T$.  We know~\cite{PS,Lo-Chau} that once Bob and
Charlie authenticate the CSS protocol the probability that
intercepts by Alice go undetected is exponentially close to
$1$. In the event the protocol is authenticated Alice's
mutual information about the key~($M_B$) is exponentially
small.  So, any attack by Alice on the communication between
Bob and Charlie cannot help her gain more than an
exponentially small amount of information about $R_B$
without going undetected with a probability exponentially
close to $1$. So with very high probability, attacks by
Alice are unsuccessful.

Consider an attack by Alice on the devices used to implement
the CSS based key distribution scheme. A CSS based scheme
can be turned into an equivalent modified BB-84
scheme~\cite{PS}, we need only analyze attacks on the latter
to discuss the security of the former.  Implementations of
QKD employ devices that may not adhere to the strict
assumptions made while proving their unconditional security.
This allows for side channels for eavesdroppers to attack.
These attacks can also be tackled. One can use
measurement-device-independent quantum key
distribution~\cite{MDIQKD} and appropriate experimental
designs~\cite{sol_1,sol_2} to achieve this.
Specifically  it has been shown that we can implement key distribution such
that it is immune to all side channel
attacks~\cite{sol_2}.

In the event the attacks are unsuccessful, then we
need only care about the noise. But as we saw
earlier the CSS protocol is robust as long as the
noise is under an acceptable level.  Since the
protocol is symmetric with respect to Alice and
Bob, any attacks by Bob are also ruled out. We
note that Charlie has access to $R_A$, $R_B$,
$C_T$ and $C$ and is restricted to be semi-honest.
It is easy to see that under these restrictions,
he can gain no information about $M_A$ or $M_B$. 


\subsection{Dishonest Third Party}
It is possible to modify our protocol to achieve
three party QPC for weaker conditions on the third
party. We allow the third party to be dis-honest,
in the sense that he may return incorrect
comparisons to Alice and Bob. We note that by
providing false results Charlie does not stand to
gain any information about the private strings of
Alice and Bob.  We adapt the technique
from~\cite{QPCs} for our purposes.  Alice and Bob,
share $m$ strings whose values are known to them.
They repeat the QPC protocol(as described above)
$m + 1$ times. They compare $m$ known strings and
$1$ secret string. Their secret strings are
compared at some random repetition, known to Alice
and Bob but unknown to Charlie. This prevents
Charlie from being dishonest. In the event Charlie
tries to give false information to Alice and Bob,
he is caught with high
probability($1-\frac{\displaystyle
1}{\displaystyle m+1}$).


\section{Conclusions}
\label{sec:conc}
We analyze EPR based three-party QPC under noisy
conditions and show that it is not robust under
any amount of bit flip noise and depolarizing
noise.  We then present a CSS based protocol that
is robust against noise and secure under general
attacks, as long as the noise is under an
acceptable rate. 

It is important to compare our work with the
available classical and quantum protocols
in the literature.  Recently a protocol
using Quantum Key Distribution(QKD)~\cite{QPCs}
have been proposed.  This protocol does not
consider noisy channels or side channel attacks.
Though it is possible from our analysis above, to
extend their work to the noisy channel case. In
terms of resources, for the case of a semi-honest
third party, their protocol achieves QPC using $4$
QKD relays each sharing $n$ bits of information.
In comparison our protocols uses $2$ QKD like
relays, decreasing the quantum resources and
communication complexity by a factor of $2$.
However, the overall communication complexity and
quantum resources(in terms of entangled states
used to implement a QKD) are still $O(n)$.

Several classical protocols have been designed to
perform two-party and multi-party secure
computation. These protocols either work under an
honest majority~\cite{rabin} or a \textit{Common
Reference String}(CRS) along with complexity
assumptions~\cite{canetti} or demand access to a
trusted dealer~\cite{damgard}~(implemented using
public key technique) but are able to tackle both
passive and active adversaries.  It is well known
that certain complexity assumptions such as
absence of polynomial time algorithms for prime
factorization or discrete logarithm are invalid
when the adversary has access to quantum
resources~\cite{gr8shor}. On the other hand it is
possible to use classical public-key cryptosystems
based on the hardness of learning with
errors~\cite{ragev}.  These cryptosystems cannot
be broken by quantum algorithms presently known to
us. Implementations of public key cryptosystems
are expensive but can be done with
$O(\text{poly}(n))$ classical resources.  In our
work we consider only $2$ parties and propose a
protocol to compute a single function~(equality)
but allow the parties to be corrupted by an
adversary who however does not inject incorrect
information into the protocol.  While we do not
need complexity assumptions we do need secure
channels between the interested parties and we
take into account the resources expended in
creating secure channels.  In our proposal the
resources~(classical and quantum) utilized to
implement the protocol from scratch are linear in
the size of the input. Our proposal based on
previous work demands a trusted third party but we
show how that assumption can be relaxed by
repeating the protocol several times, consequently
incurring a cost which is still linear in the size
of the input.

We note that our protocol no longer uses EPR
states, but requires the used of CSS code states.
In order to send CSS encoded information we may
require multiqubit channels.  In order to perform
QPC under noise we exploit the connection between
CSS codes and key distribution. This enables us to
provide unconditional security for QPC in real
time implementation schemes.

It would be interesting to see if other QPC
protocols that use $d$ level quantum systems or
Greenberger-Horne-Zeilinger (GHZ) states can also
be made unconditionally secure against all
possible attacks. It would also be  worthwhile to
explore protocols that work under milder
restrictions on the third party and protocols that
can work for multi-party and implement a wider
class of functions.

\begin{acknowledgements}
The work described above has been supported in part 
by the INSPIRE fellowship administered by the Department of 
Science \& Technology (DST), India and the National Science 
Foundation through Grant PHY-1068331. VS thanks Dan Stahlke 
and Valerio Pastro for useful discussions.
\end{acknowledgements}

\end{document}